# Pancharatnam-Berry Phase as the Origin of Vector Nature Observed in Hermite–Gaussian Superposition States

A. Srinivasa Rao[1,2,3]

[1]Graduate School of Engineering, Chiba University, Chiba 263-8522, Japan
[2]Molecular Chirality Research Centre, Chiba University, Chiba 263-8522, Japan
[3]Institute for Advanced Academic Research, Chiba University, Chiba, 263-8522, Japan
*asvrao@chiba-u.jp, sri.jsp7@gmail.com*

**Abstract**

Superposition of orthogonal Hermite–Gaussian (HG) modes in orthogonal linear polarization states is one of the techniques used for the experimental realization of lower-order optical vector beams and vector vortex lattices. To date, it has been widely believed that the vector nature arising from this technique originates from the spatial intensity distribution of the superposed HG modes. Here, we report that the vector characteristics observed during the characterization of vector modes generated via HG mode superposition arise from the azimuthally inhomogeneous intensity distribution and the polarization-dependent Pancharatnam–Berry (PB) phase. Our analytical calculations confirm that the vector nature is not inherently present in the superposition state; rather, it becomes observable due to polarization-based optical elements used in the characterization process. This insight provides a fundamental clarification of the physical origin of the experimentally observed vector nature in HG-mode superposition.

**1. Introduction**

Optical vector beams, characterized by non-uniform polarization distributions, have played a significant role in structured light applications due to their unique properties, particularly under tight focusing conditions. For reference, several prominent applications are listed here. Surface structuring of polarization-sensitive materials can be achieved by direct exposure to the non-uniform polarization distribution present across the beam cross-section of vector beams, enabling the fabrication of micro- and nano-scale structures. These structures can be systematically controlled by adjusting input parameters such as intensity, polarization, and phase [1–3]. The controlled local electric-field oscillations associated with linear polarization components have been successfully employed to manipulate particle motion in optical trapping systems [4–6]. The reduced focal spot size achieved with tightly focused radially polarized (RP) beams has been utilized in super-resolution imaging and microscale drilling [7–10]. Radial and azimuthal vector vortex modes can also mitigate beam asymmetry at interfaces arising from unequal Fresnel reflection losses of s- and p-polarized components [11]. Furthermore, these modes have been explored in optical communication systems, including data encryption schemes based on polarization [12-14].

The most well-known vector beams formed by non-uniformly distributed linear polarization states are cylindrical vector beams [15]. These beams can be generated directly from a laser cavity by properly controlling the intracavity oscillating modes [16–18], or through mode conversion of a Gaussian beam using external diffractive optical elements [19–21]. In addition to these approaches, another frequently used technique for realizing vector modes is the superposition of scalar spatial modes [22–24]. The most common implementation of this method involves the superposition of orthogonal Laguerre–Gaussian (LG) modes in the circular polarization basis. In this approach, the non-uniform polarization distribution across the beam cross-section is generally attributed to the azimuthal phase gradient. Recently, however, we demonstrated that the vector characteristics observed during the characterization of modes generated by this method arise from the Pancharatnam–Berry (PB) phase rather than from the azimuthal phase gradient alone [25]. Using a similar approach, first-order vector beams can also be generated by superposing first-order Hermite–Gaussian (HG) modes in the linear polarization basis [22], where a non-uniform azimuthal intensity distribution is typically considered responsible for the emergence of vector nature.

Here, we report the fundamental concept underlying the experimental observation of vectorial properties in superposition states formed by HG modes. In our theoretical analysis, we examine the azimuthal intensity distribution both with and without the inclusion of the polarization-dependent PB phase to investigate the origin of the experimentally observed vector nature. We show that the vector-beam signatures observed experimentally arise from polarization-dependent geometric phase effects rather than from an intrinsically spatially varying polarization distribution of the optical field. Furthermore, we extend our analysis to the superposition of higher-order HG modes, which is commonly believed to give rise to vector vortex lattice structures.

## 2. Superposition of HG modes for vector beam generation

The vector beam constructed by the superposition of orthogonal HG modes in orthogonal linear polarization states is given by

$$|V\rangle = \cos\left(\frac{\theta}{2}\right) \cdot HG_{m,n}|L(\beta)\rangle + \sin\left(\frac{\theta}{2}\right) \cdot e^{i\phi_d} \cdot HG_{m',n'}\left|L\left(\beta + \frac{\pi}{2}\right)\right\rangle. \quad (1)$$

Here, the angle $\theta$ ($0 \leq \theta \leq \pi$) represents the weighting factor in the superposition. The second angular parameter, $\phi_d$ ($0 \leq \phi_d \leq 2\pi$), denotes the dynamical phase difference and determines the relative phase between the superposed modes. The ket vector $|L(\beta)\rangle$ represents a linear polarization state oriented at an angle $\beta$. It is generally believed that, by interchanging the polarization basis vectors and controlling the relative phase between the superposed modes, various types of vector beams can be generated. Typically, the superposition of first-order HG modes $\{HG_{1,0}, HG_{0,1}\}$ is considered for the realization of lower-order vector beams, as shown in Fig. 1. However, in such a superposition state, the individual modes are present in orthogonal polarization states and therefore do not interfere. Moreover, electromagnetic waves (or photons) do not interact in free space. Consequently, there is no mechanism for generating a spatially varying vector polarization distribution across the beam cross-section during free-space propagation.

## 3. Analysis of polarization state transfer of Superposition state

The most common method for investigating the vector properties of a superposition state is through analysis of its polarization state. This analysis is typically performed using polarization optics, such as a polarizer, half-wave plate, and quarter-wave plate. However, a major limitation of this approach is that the polarization state is transformed during the measurement process, rather than being characterized in its original form and without altering the spatial modes. A straightforward and widely used technique for characterizing linearly polarized vector beams involves analyzing their polarization state using a rotating linear polarizer. In this method, when a polarizer $\hat{P}(\alpha)$, whose transmission axis is oriented at an angle $\alpha$ with respect to the horizontal, is applied to the superposition state, the resulting output state takes the form of

$$\hat{P}(\alpha)|V\rangle = \frac{1}{\sqrt{2}}\{HG_{1,0} - e^{i\phi_d} \cdot HG_{0,1} \cdot \tan(\beta - \alpha)\}|L(\alpha)\rangle. \quad (2)$$

This mathematical expression constitutes the central result of our work. The polarizer transfers the superposed modes into a single polarization state, and consequently, interference becomes observable. The resulting interference pattern is governed by two phase terms. The first is the dynamical phase, $\phi_d$. The second is the relative phase introduced between the two superposed modes by the polarizer, expressed as $\tan(\beta - \alpha)$. Here, the rotation of the polarizer by an angle $\alpha$ transports the superposition state to a position corresponding to $2\alpha$ on the Poincaré sphere. Thus, one complete rotation of the polarizer produces two rotations along the equator of the sphere. Indeed, the PB phase generated by the polarizer is equal to $\alpha$ [26–28]. This phase is incorporated into Eq. 2 through the polarizer operator. In this framework, the phase introduced by the polarizer acts as a weighting factor, since the initial and final states lie on the same trajectory on the Poincaré sphere. As an illustrative example, we consider the state transport of superpotion state correspond to RP beam on the Poincaré sphere (Fig. 2). It is important to note that both the initial and final states are superposition states. The key difference is that, in the initial state, the spatial modes are associated with orthogonal polarizations, leading to the addition of intensities. In contrast, in the final state, the same spatial modes are projected onto a single polarization state, resulting in the addition of amplitudes. The additional phase generated during this transport process is equal to the polarizer angle, $\alpha$.

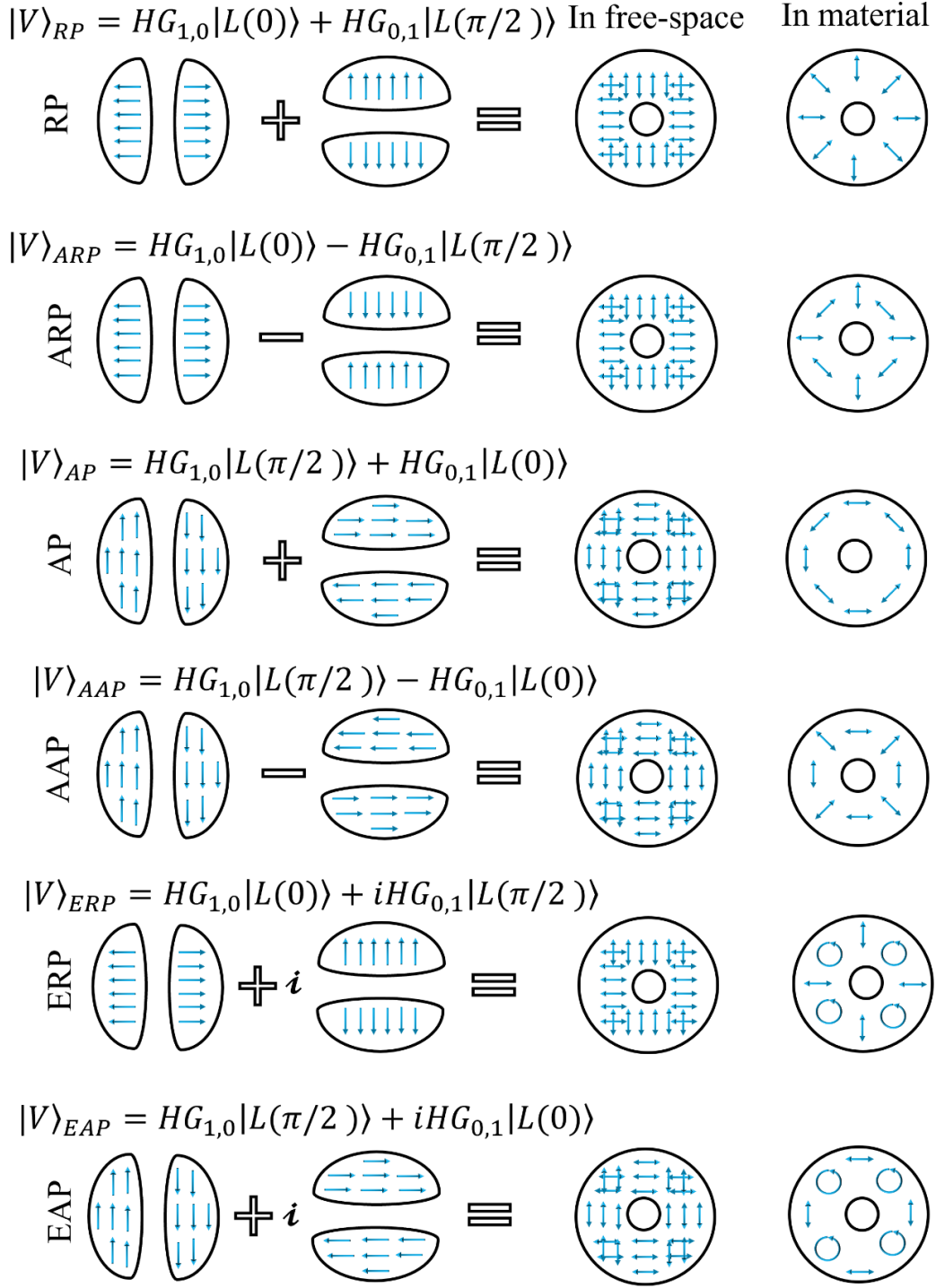

Fig. 1. Schematic illustration of first-order vector beam generation via superposition of orthogonally polarized first-order Hermite–Gaussian modes. The superposed state exhibits an annular intensity distribution with two orthogonal linear polarization components during free-space propagation (third column). The non-uniform polarization distribution across the beam cross-section is revealed using polarization-analysis optics (fourth column). RP: radial polarization; ARP: anti-radial polarization; AP: azimuthal polarization; AAP: anti-azimuthal polarization; ERP: elliptical radial polarization; EAP: elliptical azimuthal polarization.

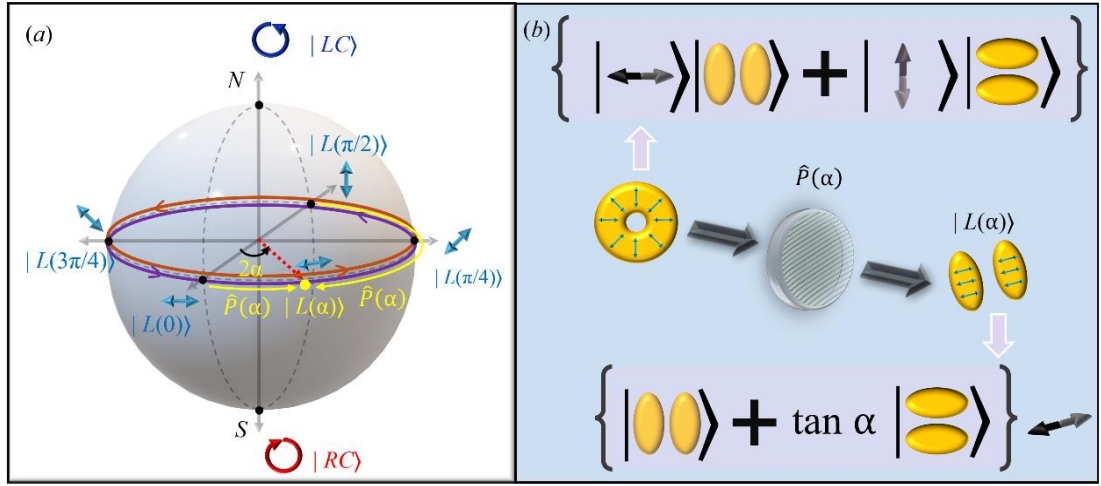

Fig. 2. Evolution of the geometric phase between superposed Hermite–Gaussian modes on the polarization Poincaré sphere during polarization-state transfer of the superposition state, assumed to form a radial polarization (RP) vector beam, through a linear polarizer. (*a*) The superposition state is projected onto a single polarization state $|L(\alpha)\rangle$ located on the equator of the polarization Poincaré sphere. (*b*) Polarization-state transfer of the superposition state through a linear polarizer whose transmission axis is oriented at an angle $\alpha$.

To intrinsically investigate the rotating lobe structure generated by the polarizer as its transmission axis rotates, we consider two key factors: the inhomogeneous intensity distribution of the HG modes and the polarization-dependent PB phase introduced by the polarizer. First, we analyze the transmitted intensity through the polarizer without taking the PB phase into account. Subsequently, we incorporate the PB phase and examine the resulting modifications in the intensity distribution of the output modes.

### 3.1. Investigation of rotating lobe structures without geometric phase

The superposition state with a doughnut-shaped intensity profile transforms into a rotatable first-order HG mode after passing through the polarizer (Fig. 3). To determine the relative intensity effect introduced by the polarizer, we initially consider only the transmitted intensity of the individual modes contributing to the interference, while neglecting the PB phase generated by the polarizer. When the transmission axis of the polarizer is oriented horizontally, it transmits the horizontally polarized HG mode and rejects the vertically polarized HG mode. A similar situation occurs when the polarizer axis is aligned vertically. In both of these configurations, no interference occurs between the spatial modes. However, when the transmission axis is oriented at any angle other than these two cases, interference is produced between the two spatial modes. Although the spatial intensity distributions of the superposed modes do not perfectly overlap, interference still occurs with visibility $\eta = 1$. For dynamical phase differences $\phi_d = 0$ and $\phi_d = \pi$, the interference pattern exhibits a lobe structure. The angular position of the lobe structure depends on the relative amplitudes of the superposed modes. Furthermore, when the superposed states interfere with equal weight factors—corresponding to diagonal and anti-diagonal orientations of the polarizer axis—the resulting field exhibits a vortex phase with an annular intensity distribution for $\phi_d = \pi/2$ and $\phi_d = 3\pi/2$. For a fixed dynamical phase difference, only a single vortex phase is obtained; however, the phase gradients have opposite signs for $\phi_d = \pi/2$ and $\phi_d = 3\pi/2$. In contrast, unequal weight factors at the same dynamical phase difference lead to the formation of a Hermite–Laguerre–Gaussian (HLG) beam [29], which occurs when the polarizer axis is oriented at angles other than horizontal, vertical, diagonal, or anti-diagonal. It is worth noting that, for a given dynamical phase, the rotation or modification of the lobe structure is confined within $\pi/2$ radians and remains within a single quadrant of the Cartesian coordinate system. From this observation, we infer that complete rotation of the lobe structure cannot be achieved solely by varying the relative amplitudes of the spatial modes in the interference process for a fixed dynamical phase difference.

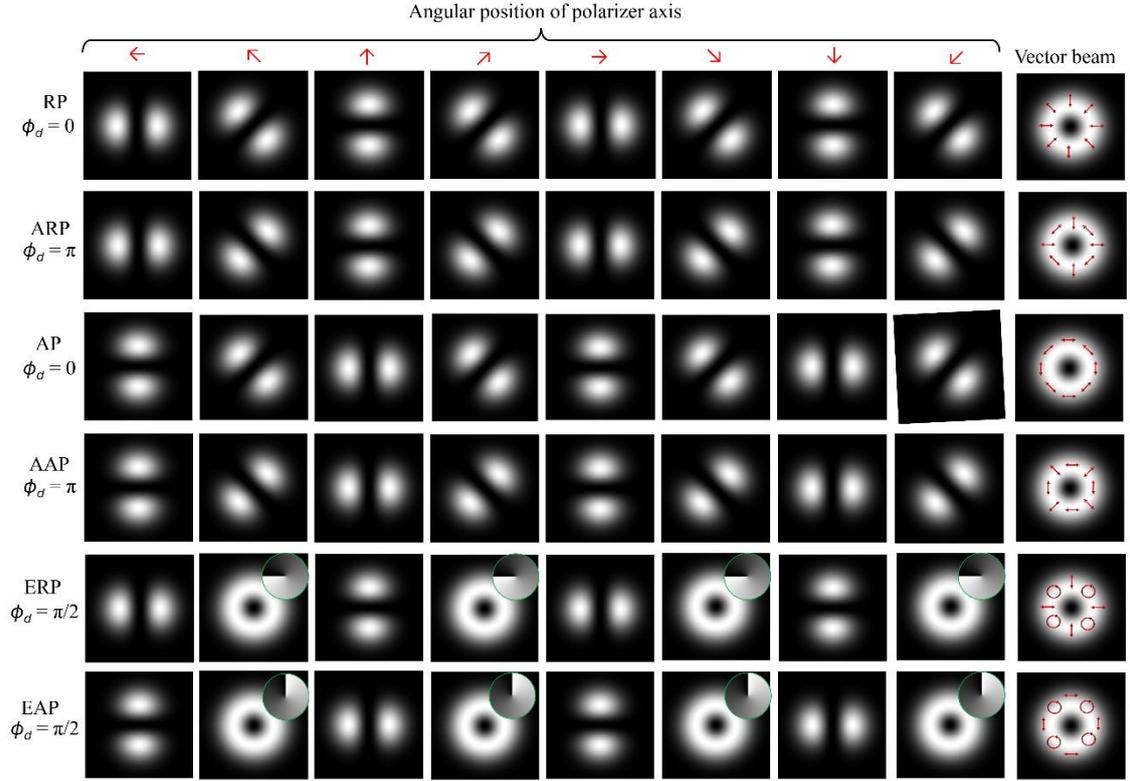

Fig. 3. Superposition states after transmission through a linear polarizer oriented at different angular positions, calculated without including the Pancharatnam–Berry phase induced by the polarizer. The corresponding vector beams are shown on the right for reference (RP: radial polarization; ARP: anti-radial polarization; AP: azimuthal polarization; AAP: anti-azimuthal polarization; ERP: elliptical radial polarization; EAP: elliptical azimuthal polarization). The red arrows at the top indicate the orientation of the polarizer transmission axis.

### 3.2. Investigation of rotating lobe structures with geometric phase

We repeat all the conditions used to generate the spatial intensity profiles shown in Fig. 3, now incorporating the PB phase. The corresponding results are presented in Fig. 4. The superposition state composed of the $HG_{1,0}$ mode in horizontal polarization and the $HG_{0,1}$ mode in vertical polarization at $\phi_d = 0$ produces a fully rotating lobe structure that follows the RP vector nature. The same state at $\phi_d = \pi$ generates a rotating lobe structure consistent with the anti-radially polarized (ARP) vector nature. Similarly, the superposition state formed by the $HG_{0,1}$ mode in horizontal polarization and the $HG_{1,0}$ mode in vertical polarization at $\phi_d = 0$ produces a rotating lobe structure corresponding to the azimuthally polarized (AP) vector nature. When $\phi_d = \pi$, this state results in a rotating lobe structure exhibiting anti-azimuthally polarized (AAP) vector characteristics. The same behavior is observed for the remaining two superposition cases, corresponding to elliptically radially polarized (ERP) and elliptically azimuthally polarized (EAP) vector beams. From these results, we infer that the complete rotation of the lobe structure which underlies our interpretation of the HG mode superposition as a vector beam originates from the PB phase introduced by the polarizer.

For a dynamical phase difference of $\phi_d = \pi/2$, the superposition state produces identical optical vortices when the polarizer transmission axis is oriented at $\alpha = \pi/4$ and $\alpha = 5\pi/4$. Similarly, identical vortices are obtained when the transmission axis is set to $\alpha = 3\pi/4$ and $\alpha = 7\pi/4$. Notably, the vortices generated at the diagonal and anti-diagonal orientations of the polarizer possess opposite handedness. Thus, the superposition state can produce optical vortices with either left- or right-handed helicity simply by rotating the polarizer axis between diagonal and anti-diagonal positions. This switching occurs at the expense of a 50% power loss due to projection, but without requiring any modification to the experimental alignment. Such a capability may offer practical advantages in applications involving optical vortices.

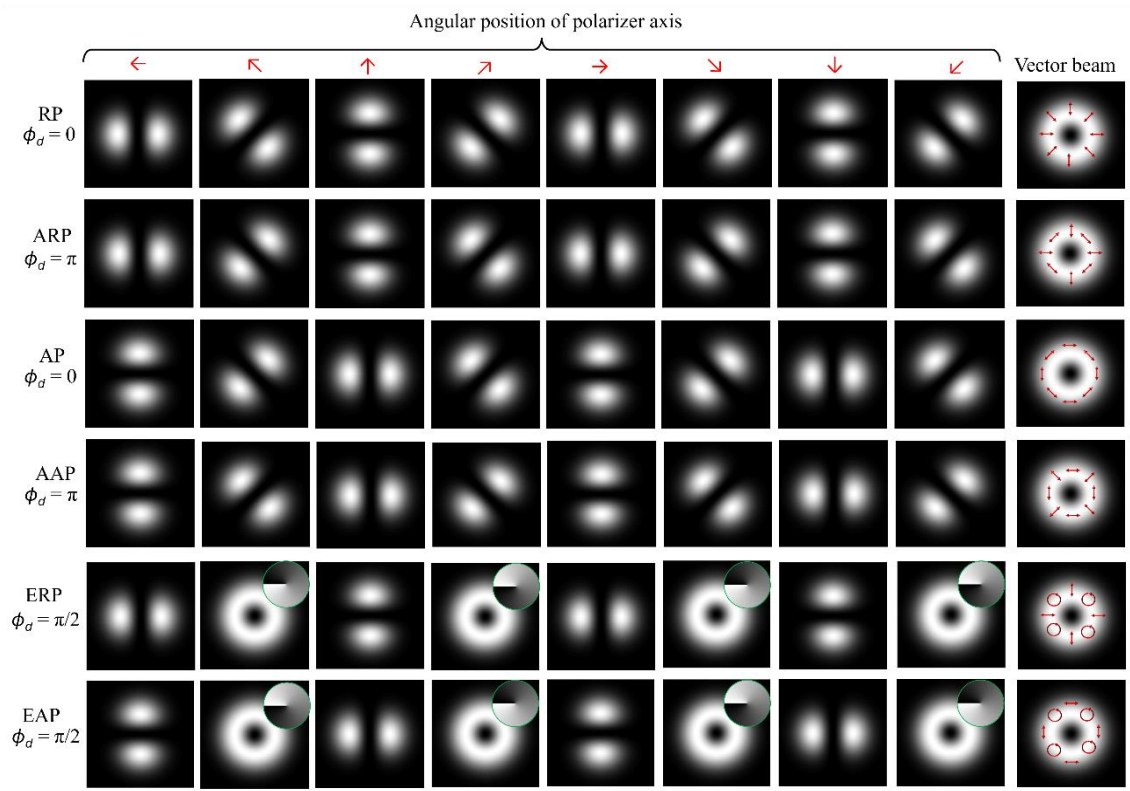

Fig. 4. Superposition states after transmission through a linear polarizer oriented at different angular positions, including the Pancharatnam–Berry phase induced by the polarizer. The corresponding vector beams are shown on the right for reference (RP: radial polarization; ARP: anti-radial polarization; AP: azimuthal polarization; AAP: anti-azimuthal polarization; ERP: elliptical radial polarization; EAP: elliptical azimuthal polarization). The red arrows at the top indicate the orientation of the polarizer transmission axis.

## 4. Optical lattice structures with superposition of higher order HG modes

In the previous section, we considered the superposition of lower-order orthogonal HG modes, which resulted in lobe structures and first-order optical vortices. In a similar manner, the superposition of higher-order HG modes within the same polarization state produces optical vortex lattice textures containing multiple phase singularities located at intensity nulls (top two rows of Fig. 5) [30–32]. Interestingly, a similar intensity distribution can also be obtained by simply adding the intensities of the superposed modes when they are prepared in orthogonal polarization states (bottom row of Fig. 5). Although this configuration resembles a vortex lattice structure, it does not genuinely constitute one. From this observation, we infer that a stable lattice of intensity nulls does not necessarily imply the presence of phase singularities.

The superposition of higher-order HG modes prepared in orthogonal polarization states is often regarded as a vector vortex lattice structure [33]. When such a superposition state is transmitted through a rotating polarizer, the output intensity becomes angle-dependent due to both the transmission of the individual spatial modes and their mutual interference. As an example, consider a superposition state formed by the $HG_{4,0}$ and $HG_{0,4}$ modes in orthogonal linear polarization states with dynamical phase difference $\phi_d = \pi/2$. When this state is passed through a polarizer whose transmission axis is rotated to different orientations (Fig. 6), the transmitted intensity distribution is completely governed by the polarizer angle. In all cases, the output mode is linearly polarized and therefore represents a scalar field. When the polarizer transmission axis is oriented along the diagonal or anti-diagonal directions, a scalar vortex lattice structure is generated. Notably, the phase singularities formed at the diagonal orientation have opposite topological charge compared to those formed at the anti-diagonal orientation. A key advantage of this type of superposition state is that the sign of the phase singularities in the optical vortex lattice can be switched simply by rotating the polarizer, without altering the experimental configuration. It is also worth noting that HG modes with equal mode number produce well-defined eigen vortex lattice structures, which may play a significant role in applications such as optical communications and light–matter interactions. In general, the number of phase singularities generated by a superposition state of the form $\left(\mathrm{HG}_{m,n}+i\,\mathrm{HG}_{m',n'}\right)|H\rangle$ is given by $mn' + m'n$.

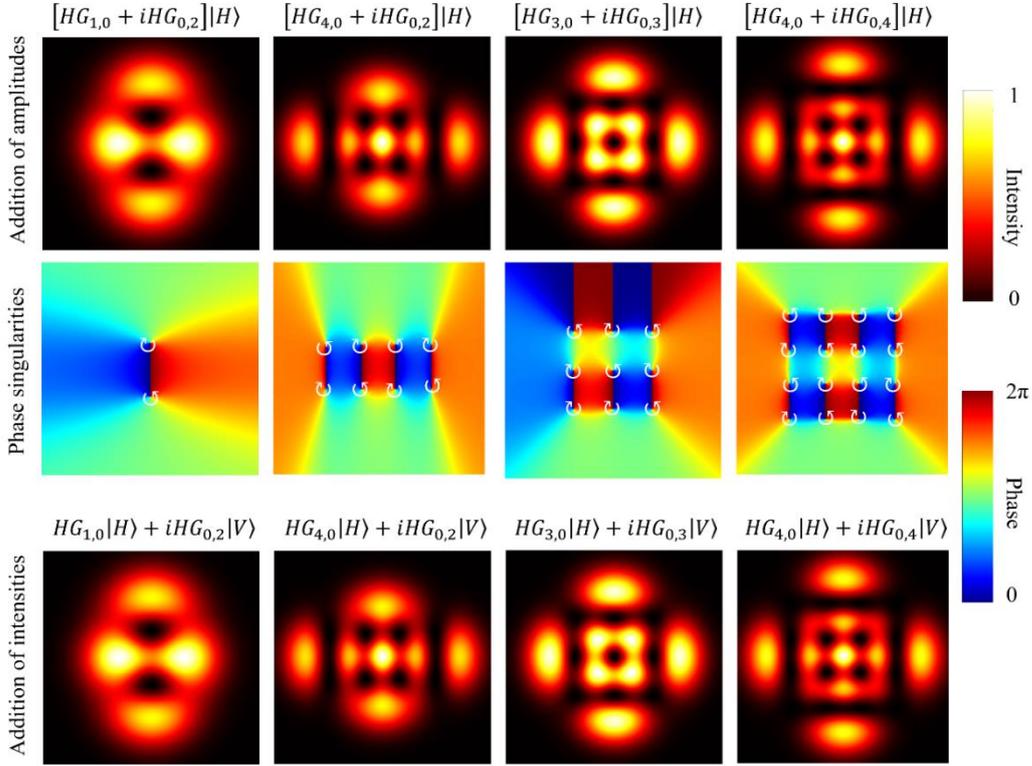

Fig. 5. Interference of higher-order orthogonal Hermite–Gaussian modes in linear polarization states. Row 1: Optical vortex lattice formed by the coherent superposition of higher-order Hermite–Gaussian modes in the same polarization state, with a dynamical phase difference of $\phi_d = \pi/2$. Row 2: Corresponding phase singularity distributions of the lattice structures shown in Row 1. Arrows at the singularity points indicate the direction of phase increase. Row 3: Superposition of higher-order Hermite–Gaussian modes in orthogonal polarization states, resulting in intensity addition without interference. Here, $|H\rangle$ denotes the horizontal polarization state $|L(0)\rangle$, and $|V\rangle$ denotes the vertical polarization state $|L(\pi/2)\rangle$.

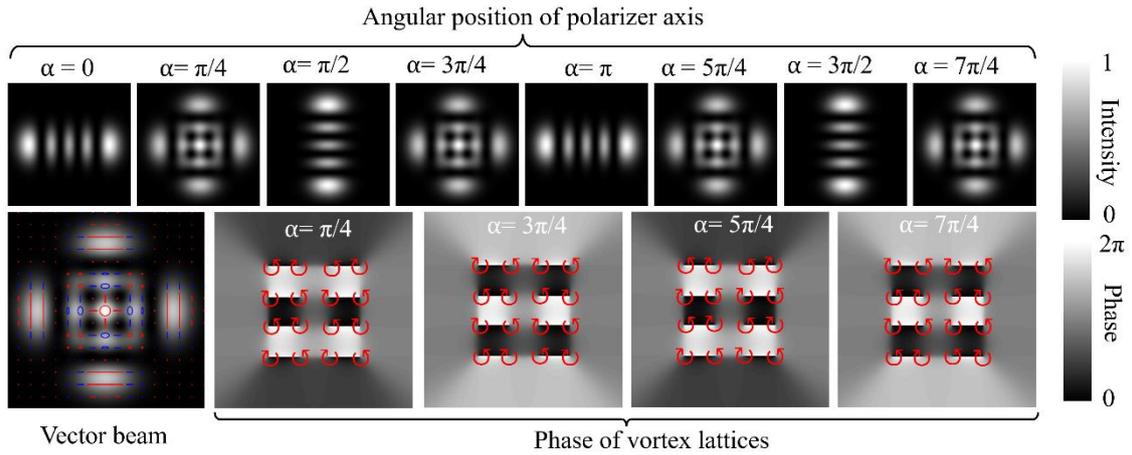

Fig. 6. Investigation of scalar and vector topological textures of the superposition state $[\mathrm{HG}_{4,0} + i\,\mathrm{HG}_{0,4}]|H\rangle$. Row 1: The superposition state formed by the $\mathrm{HG}_{4,0}$ and $\mathrm{HG}_{0,4}$ modes in orthogonal linear polarization states is transmitted through a linear polarizer whose transmission axis is rotated to different angular orientations. Row 2: The first image shows the expected optical vector vortex lattice associated with the superposition state. Red ellipses denote right-handed polarization states, and blue ellipses denote left-handed polarization states. The remaining images correspond to scalar phase lattice structures obtained when the polarizer transmission axis is oriented along the diagonal and anti-diagonal directions. Red circular arrows at the phase singularities indicate the direction of azimuthal phase increase.

## 5. Conclusion

The experimentally observed vector nature of superposed orthogonal HG modes prepared in the linear polarization basis arises from two primary factors: the inhomogeneous azimuthal intensity distribution of the HG modes and the polarization-dependent PB phase. In the present study, we employed a polarizer as an optical element to probe the superposition state and to verify the apparent vector characteristics arising from both the intensity distribution and the PB phase. Our analysis was further extended to optical vector vortex lattice structures. Based on these investigations, we conclude that the vector modes realized through the superposition technique do not constitute true vector beams in a strict sense. Rather, they exhibit apparent vector characteristics that originate from artefacts associated with the measurement process.